# All-optical determination of one or two emitters using quantum polarization with nitrogen-vacancy centers in diamond

Davin Yue Ming Peng,[1] Josef G. Worboys,[1] Qiang Sun,[1] Shuo Li,[1] Marco Capelli,[1] Shinobu Onoda,[2] Takeshi Ohshima,[2] Philipp Reineck,[1] Brant C. Gibson,[1] and Andrew D. Greentree[1]

[1]*Australian Research Council Centre of Excellence for Nanoscale BioPhotonics, RMIT University, Melbourne, VIC 3001, Australia*

[2]*Takasaki Advanced Radiation Research Institute, National Institutes for Quantum Science and Technology,1233 Watanuki, Takasaki, 370-1292, Gunma, Japan*

Abstract

Qubit technologies using nitrogen-vacancy color centers in diamonds require precise knowledge of the centers, including the number of emitters within a diffraction-limited spot and their orientations. However, the number of emitters is challenging to determine when there is finite background, which affects the precision of resulting quantum protocols. Here we show the photoluminescence (PL) intensity and quantum correlation (Hanbury Brown and Twiss) measurements as a function of polarization for one- and two-emitter systems. The sample was made by implanting low concentrations of adenine ($C_5H_5N_5$) into a low nitrogen chemical vapor deposition diamond. This approach yielded well-spaced regions with few nitrogen-vacancy centers. By mapping the PL intensity and quantum correlation as a function of polarization, we can distinguish two emitter systems from single emitters with background, providing a method to quantify the background signal at implanted sites, which might be different from off-site background levels. This approach also provides a valuable new all-optical mechanism for the determination of one or two emitter systems useful for quantum sensing, communication, and computation tasks.

Introduction

Quantum technologies such as quantum sensing, quantum communication, and quantum computation are expected to lead to wide-ranging opportunities, with at least as significant an impact as computers and the internet have had on society [1-4]. However, for quantum technologies to become ubiquitous requires robust and room-temperature platforms. Optically active defects in wide-bandgap semiconductors are therefore of great interest, with the nitrogen-vacancy color center in diamond being the most widely explored [5,6], although there are now emerging alternatives [7-9].

In developing practical quantum sensors, one of the critical issues that need to be addressed is characterization. In particular, for many applications, it is essential to confirm that there is one and only one emitter in a given diffraction-limited spot. Confirmation of single emitters is generally a complex problem, especially in the presence of significant background signals. Lacking robust methods to independently characterize the properties of a quantum system therefore significantly hampers efforts to optimize and extend processes.

Here we show an all-optical method for characterizing the number and orientation of one or two emitters in a diamond sample with the presence of finite background. Our method uses the Hanbury Brown and Twiss (HBT) antibunching signal obtained as a function of polarization. By considering the polarization-dependent intensity and HBT signal for each emitter with the unpolarized background, we are able to determine the number of emitters and their orientation. These results also confirm the theoretical treatment of two-emitter HBT presented in Worboys *et al.* [10]. We illustrate our method using nitrogen-vacancy (NV) color centers in diamond, created by implanting low concentrations of adenine ($C_5H_5N_5$) into low nitrogen chemical vapor deposition (CVD) grown diamond. Although, magnetic resonance methods [11] have been used to identify the number and calculate the separation between closely located NV centers through magnetic field measurements of NV-NV coupling. Here we show we can distinguish between sites with one and two emitters using a full optical method.

A standard method to identify single photon emitters is through the use of the HBT setup [12,13]. The HBT experiment measures the correlations between photon arrival times using two single-photon detectors, as shown in Fig.1(a). In addition to the HBT setup, we introduce a single linear polarizer before the detector, as displayed in Fig. 1(b). By rotating the linear polarizer we are able to vary the photon collection probability from each NV center. Fig. 1(c) shows $g^{(2)}(0)$ as a function of polarizer angle for obtained antibunching signal as a function of emission polarization angle. The polarization dependence of the NV center PL arises from deexcitation via a superposition of one of two orthogonal dipole moments perpendicular to the N-V axis, as shown in Fig.1(b) [14-16] (See Supplemental Material Fig. S1 and S2 for more details [17]). While the NV PL properties also depend on the polarization of the excitation beam (see Supplemental Material Fig. S3 [17]), here we will focus on the polarization of the emitted light. PL intensity measurements with varying polarization angles of the excitation beam have been reported in [18] to characterize the angular-dependent of NV centers in nanodiamonds. A method for the determination of the orientation of NV centers using radially polarized excitation beams was reported by Dolan *et al.* [19].

The coincidence count of photon arrival time delay, $\tau$ (ns), between two detectors is expressed by the second-order correlation function $g^{(2)}(\tau)$, where the coincidence counts are normalized with uncorrelated coincidence counts at time $\tau = \pm\infty$ shown in Fig. 1(d). The antibunching signal, $g^{(2)}(0)$, is primarily of interest as it offers the most information regarding the number of single photon emitters (SPEs) present within the region of interest [20].

The general form of $g^{(2)}(0)$ for *n* SPEs of equal brightness is

$$g_n^{(2)}(0) = \left(1 - \frac{1}{n}\right)$$

(1)

In particular, the value of $g_n^{(2)}(0) = 0.5$ is achieved from two equal brightness emitters with no background signal. This limit is also often erroneously used to imply that any value of $g^{(2)}(0)$ less than 0.5 is associated with an SPE and a background signal. However, two unequal brightness emitters will also give rise to $g^{(2)}(0) < 0.5$ [10].

Here, we demonstrate control of $g_n^{(2)}(0)$ for two emitters of unequal PL intensity in the presence of a finite background signal. The control is achieved by varying the orientation of a polarizer inserted between the sample and the detectors as shown in Fig. 1(b). Our method allows characterization of the number of emitters, their orientation within a sample, and the background level. This paper is organized as follows. First, we describe the dipole emission properties of single NV color centers in diamond. We then examine the intensity and HBT signals obtained for two-NV systems as a function of emission polarization angle.

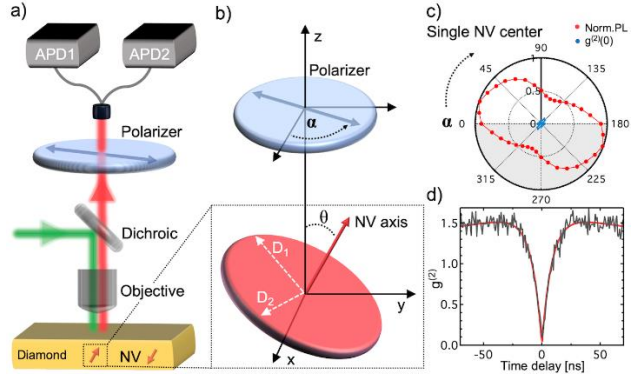

**Figure 1.** (a) Schematic of the custom-built confocal fluorescence microscope including a linear polarizer and a Hanbury Brown and Twiss setup. NV centers are excited with a 532 nm laser (green line), and the PL signal (red line) is separated from the excitation beam with a dichroic mirror. A linear polarizer is rotated to vary the relative brightness of centers. Coincidences between the detectors APD1 and APD2 are monitored using a time-correlated single photon counting module. (b) Schematic indicating the double dipole model of the NV center. The NV center can relax via a superposition of dipoles D1 or D2, leading to the characteristic intensity pattern measurement with the linear polarizer. (c) Measured $g^{(2)}(0)$, blue dots, and normalized PL, red dots, as a function of polarization angle in degrees for a single NV center. Photoluminescence shows the characteristic 'peanut' shape, whereas $g^{(2)}(0)$ is unchanged against the background for a single emitter. (d) Second order correlation $g^{(2)}$ measurement of a single NV center with $g^{(2)}(0)$ value of 0.03.

The value of $g^{(2)}(0)$ for two SPEs with unequal brightness is [10],

$$g_2^{(2)}(0) = \frac{2P_1P_2}{(P_1+P_2)^2} = \frac{2\alpha}{(1+\alpha)^2}$$

(2)

where $\alpha = P_1/P_2$ is the ratio of the probability of detecting a photon from SPE 1 ($P_1$) and 2 ($P_2$).

This result shows that the maximum value of $g^{(2)}(0) = 0.5$ is only achieved for the equal brightness case of $P_1 = P_2$. As any other brightness ratio leads to values less than 0.5, and background can also contribute to non-zero $g^{(2)}(0)$ for a single emitter, discriminating between these cases (more than one emitter and non-zero background) is nontrivial and an important goal of this work. To consider the effect of the background, we treat the background as coming from a large number of weak single-photon emitters, where there is assumed to be no dependence on polarization for the background. In this limit, the probability that any one of the background emitters gives a photon is small, but the product of the number of background emitters with the probability of detecting a photon from one emitter is not negligible. In this case, $g^{(2)}(0)$ is

$$g_2^{(2)}(0) = \frac{2P_1P_2 + 2(P_1 + P_2)NP_\gamma + (NP_\gamma)^2}{(P_1 + P_2 + NP_\gamma)^2}$$

(3)

where $N \gg 1$ is the number of background emitters, which have equal detection probability, $P_\gamma \ll P_1, P_2$, but $NP_\gamma \sim P_1, P_2$.

We investigated NV centers in a single crystal diamond at room temperature using a custom-built confocal microscope with an incorporated HBT setup (Fig.1 (a) and see Fig S4 in the Supplemental Material for details [17]). The sample studied is an electronic grade [100] crystal orientation single-crystal diamond implanted with adenine ($C_5H_5N_5$). This implantation strategy yields the proportion of single, double, and triple NV centers of 22.3%, 3.4%, and 0.1%, respectively. The sample was implanted with 70 keV-$C_5H_5N_5$ ions with the fluence of $1 \times 10^8$ ions/cm$^2$ at room temperature following the method detailed in Haruyama et al. [21]. Before and after implantation, the sample was acid cleaned in a mixture of sulfuric and nitric acid (ratio 1:3) heated to ~200 °C for 0.5 hours to remove surface contamination. The sample was then annealed at 1000 °C for 2 hours in a vacuum to combine nitrogen with vacancies. After annealing, the surface treatment was as follows: (1) the sample was cleaned in the above acid mixture again, and (2) the sample was annealed at 460 °C in $O_2$ ambient to oxygen terminate the surface, which has the effect of stabilizing the negative charge state of the NV center [22]. NV centers were excited with a 532 nm continuous wavelength laser. NV PL was collected with a 100× oil immersion objective and separated from the excitation signal using a 532 nm dichroic and a 697/75 nm bandpass filter. A PL map of the investigated sample and statistical analysis of the fluorescence intensity of 362 regions of interest on this map are shown in Fig. S5 in the Supplemental Material [17]. All regions of interest showed an NV fluorescence spectrum (Supplemental Material Fig. S6 [17]).

We start by performing emission polarization-dependent PL intensity and $g^{(2)}(\tau)$ measurements on PL spots within the sample and located a region of interest with a single NV center as control. Radiative NV decay occurs via a superposition of two dipoles, leading to a characteristic 'peanut' shape of PL intensity as a function of polarization [23,24], as shown in Fig. 1(c). All polarization measurements were performed with 10° polarization angle increments from the arbitrary 0° to 180°. We have chosen to mirror the data from 180° to 360° for clarity. The PL data confirms the expected polarization dependence and partially polarized nature of the NV center emission. Ideally, for a single emitter, $g^{(2)}(0) = 0$. However, background photons from the region of interest and instrument noise from detectors and electronic devices can contribute to non-zero $g^{(2)}(0)$. Because the background does not have any appreciable symmetry, we expect the background photons to exhibit no net polarization. In this case, we found $g^{(2)}(0) < 0.069$ for all emission polarization angles, as shown in Fig.1(c).

Based on the polarization dependence of the PL intensity (I) of a single NV center and equation (3), we developed a protocol to distinguish between one and two NV centers within a diffraction-limited spot in a sample with known crystal orientation. This is achieved by, first, measuring the PL intensity and $g^{(2)}(0)$ as a function of emission polarization angle and then performing a least square analysis of the experiment data against the theoretical simulation all ten of the two-emitter orientation combinations of the known crystal lattice, [100] oriented in this case.

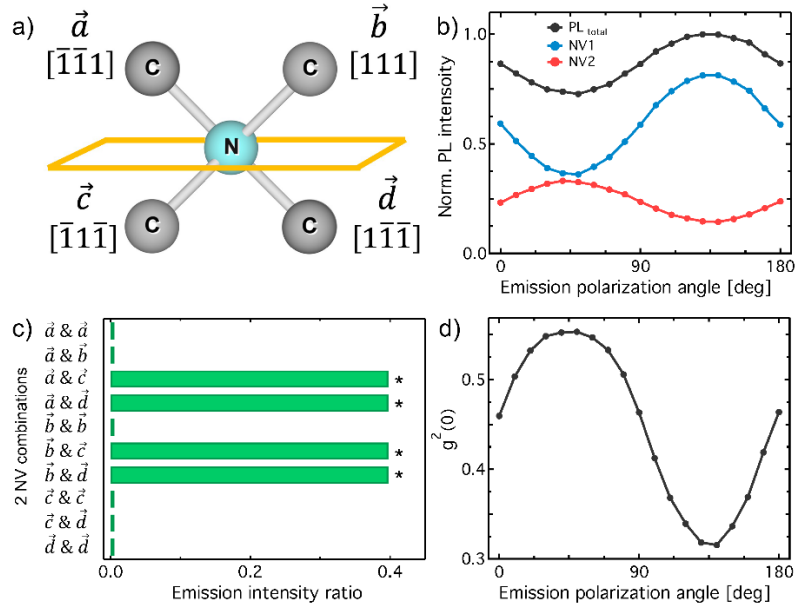

**Figure 2.** (a) Schematic of the atomic structure showing all four possible NV orientations in an [100] oriented single crystal diamond. Annotations $\vec{a}$, $\vec{b}$, $\vec{c}$ and $\vec{d}$ correspond to the respective vector coordinates used in theoretical simulations. (b) PL intensity simulation of a two NV center system as a function of emission polarization angle with NV center orientations $\vec{a}$ and $\vec{c}$ assigned in Fig. 2 (a), PL intensity (I) ratio $I_{\vec{c}}:I_{\vec{a}} = 0.4$, and background $\gamma$ of $I_{bg}: \max(I_{\vec{a}}) = 0.05$. (c) Result of intrinsic PL intensity ratio (green bar) and orientations of the two NV centers corresponding to minimum $\chi^2$ value highlighted with (*) identified using the $\chi^2$ optimization protocol. (d) $g^{(2)}(0)$ value calculated using equation (3) and PL simulation data in Fig. 2 (b).

**Theory and simulations**

The protocol was tested on the simulation data sets of two NV center systems with predetermined ground truth parameters to verify performance and reliability.

Individual NV center's emission PL intensity was simulated by the two orthogonal electric dipoles which represent the emission plane perpendicular to the NV axis. As derived in SI, the orientations of the two dipoles are not relevant if they are orthogonal to each other. The photon collector is represented by a circular objective with NA=1.98. To take the effect of the linear polarizer into account, the electric field radiated from the two dipoles is projected along the polarizer with respect to angle $\theta$. In the simulations, the polarizer is assumed to be an ideal one with a parallel transmittance of 1. Then, the emission PL intensity is the square of the magnitude of that projected electric field.

The protocol was then performed on the simulated data set, the results were analyzed and compared with the known properties of the simulated NV center system. There are four possible orientations for NV centres in the diamond lattice. We denote these orientations $\vec{a}$ to $\vec{d}$ as illustrated in Figure 2(a).

The simulated data for two NV centres with orientations $\vec{a}$ and $\vec{c}$ has an intrinsic PL intensity ratio of 0.40, $\gamma$ of 0.05, and an acquisition time (t) of 1000, where $E_n(\theta) \times t$ represents the expected number of photon arrivals. The acquisition time was introduced using POISSRND function in MATLAB to generate a random variable,

$$C_n = \text{POISSRND}(D_n(\theta) \times t)$$

(4)

where $C$ is the total counts collected from NV centre $n$, and $D$ is the photon detection rate calculated using the emission rate, NA, losses of the collection system and polarization (Supplemental Material Equation S5 [17]).

The optimization compares experimental data against simulation data for all possible combinations of NV center orientations (Fig. 2 (c)) for a two-emitter system. Our approach uses maximum likelihood estimation, minimizing the $\chi^2$ value,

$$\chi^2 = \sum_{\theta=0}^{\pi} \frac{\left(I_{Exp} - I_{\vec{a}+\vec{c}}\right)^2}{I_{\vec{a}+\vec{c}}} + \frac{\left(g^2_{Exp}(0) - g^2_{\vec{a}+\vec{c}}(0)\right)^2}{g^2_{\vec{a}+\vec{c}}(0)}$$

(5)

which is the sum of the squared differences between the fitted and experimental data for of PL intensity and g$^{(2)}$(0) as a function of polarization angle. This allows us to assess the goodness of fit for each NV orientation combination.

The capacity of the $\chi^2$ optimization protocol to identify the correct number and orientation of NV centers is demonstrated in Figure 2 (c) where the results indicated instead of the presence of a single there are two NV centers within the region of interest.

Figure 2 (c) shows the PL intensity ratio and $\chi^2$ for all possible combinations and orientations of two NV centers as determined by the $\chi^2$ optimization protocol. Here, the possibility of the presence of only one emitter is included since the relative PL intensity of the two emitters is an optimization parameter that can approach zero. Only four NV combinations ($\vec{a}\&\vec{c}$, $\vec{a}\&\vec{d}$, $\vec{b}\&\vec{c}$, and $\vec{b}\&\vec{d}$) result in $\chi^2$ values below $3 \times 10^{-4}$ indicating good quantitative agreement between the simulated data and the optimization results. All other combinations result in $\chi^2$ values of 0.21. Hence, the protocol has identified the emitter orientations of $\vec{a}$ and $\vec{c}$ as well as degenerate orientation combinations that are expected to yield identical experimental results. In the case of the [100] diamond crystal investigated here, these four NV combinations exhibit the same PL intensity polarization dependence due to their symmetry within the atomic structure with respect to the [001] optical axis of the experiment setup resulting in undistinguishable two-NV combinations for orientations $\vec{a}\&\vec{c}$, $\vec{a}\&\vec{d}$, $\vec{b}\&\vec{c}$, and $\vec{b}\&\vec{d}$.

The protocol correctly determines the intrinsic PL intensity ratio of the two emitters as 0.40. Importantly, it also shows that a fit based on a single emitter (e.g., the presence of two aligned NV centers, for example two centres in the $\vec{a}$ direction, with one center's PL intensity close to zero) yields a very poor agreement indicated by $\chi^2$ of 0.21 shown in Figure 2(c).

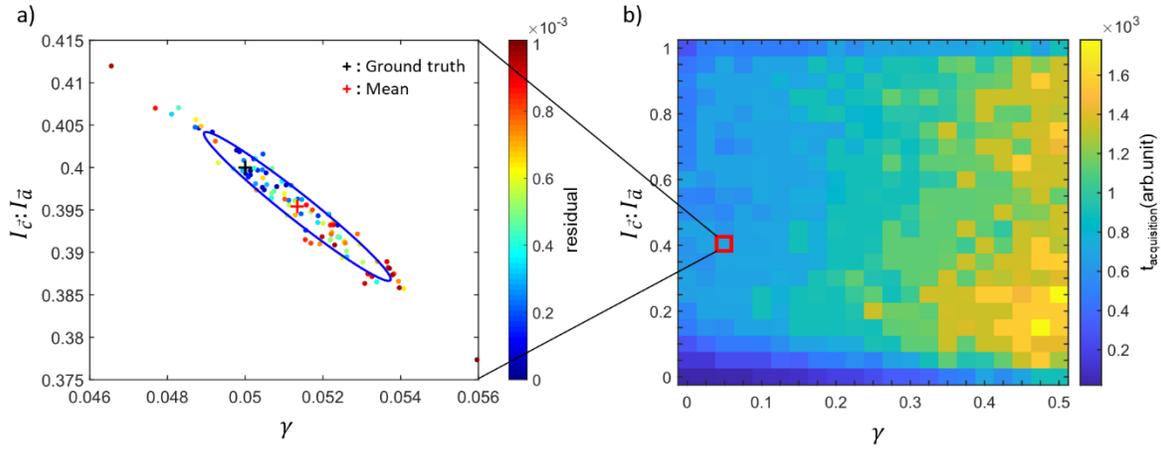

**Figure 3.** (a) Result of optimization protocol performed on 100 simulation data sets of the same ground truth parameters, PL intensity ratio $I_{\vec{c}}:I_{\vec{a}} = 0.4$ and background $\gamma$ of $I_{bg}:\max(I_{\vec{a}}) = 0.05$, plotted in the background and the PL intensity ratio with color scale representing the $\chi^2$ value. The blue oval indicates the 1σ (68.3%) confidence interval of the optimization results, the red marker shows the average result for the background and the PL intensity ratio. (b) Minimum t (acquisition) to obtain a 1σ (68.3%) confidence interval area less than an uncertainty value of ± 0.01 for both parameters $\gamma$ and $I_{\vec{c}}:I_{\vec{a}}$, plotted with respect to simulation data sets generated of corresponding value of $0 \leq \gamma \leq 0.5$ and $0 \leq I_{\vec{c}}:I_{\vec{a}} \leq 1$. This result shows low PL intensity ratio and $\gamma$ is easier to distinguish than similar PL intensity between the two NV centers.

The convergence behavior of the protocol was investigated, the $\chi^2$ optimization protocol was performed on 100 simulation data sets of the same ground truth parameters with an acquisition time t = 1000, background value and the PL intensity ratio, of one set of simulation data is plotted in Figure 3(a) with a standard deviation confidence interval of the optimization results indicated with a blue oval. The result shows a convergence on the ground truth parameters, $\gamma = 0.05$ and the PL intensity ratio of 0.40.

The same test was then performed on 441 simulation data sets with varying PL intensity ratio and $\gamma$ value parameters to determine the minimum acquisition time ($t$) required to obtain a result of 1σ (68.3%) standard deviation with a ± 0.01 uncertainty error from the ground truth value. The test result is shown in Figure 3(b), where $\gamma$ was plotted against the PL intensity ratio, and the color axis represents the minimum acquisition time required to reach the target error level determined by the one standard deviation confidence interval area.

Results in Figure 3(b) show the protocol can reach the target error level with variation in acquisition time for $\gamma$ from 0 to 0.5 and corresponding PL intensity ratio from 0 to 1. Acquisition time to reach target error increases as PL intensity ratio increases for $\gamma$ less than 0.5.

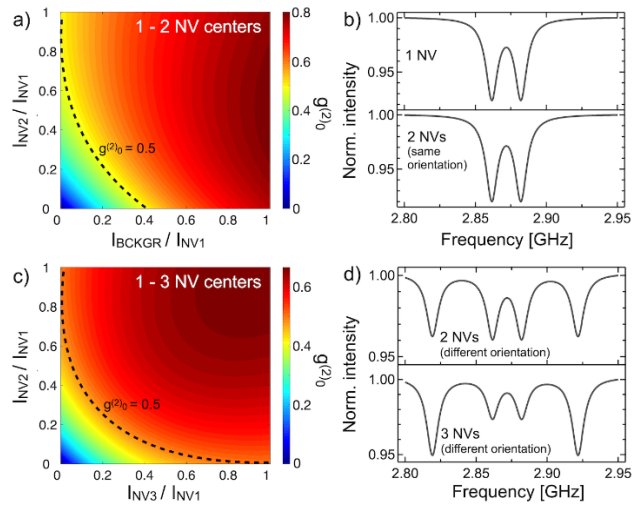

**Figure 4.** (a) Simulation of unpolarized $g^{(2)}(0)$ measurements (color scale) of a two-NV center system within a diffraction-limited spot with varying PL intensity ratio between the two NV centers and background. The black dashed line represents the contour of $g^{(2)}(0) = 0.5$. (b) Simulation of an ODMR measurement for one and two collocated NV centers of the same orientation within a diffraction-limited spot in an [100] oriented diamond crystal. Without photon counting approaches, it is typically impossible to practically distinguish the one and two-emitter cases. (c) Simulation of unpolarized $g^{(2)}(0)$ measurement (color scale) of a collocated three-NV center system, with varying brightness ratio among the three NV centers and zero background contribution. (d) Simulation of ODMR measurement for 2 ($[\bar{1}\bar{1}1]$ and $[\bar{1}1\bar{1}]$) and 3 ($[\bar{1}\bar{1}1]$, $[\bar{1}1\bar{1}]$ and $[\bar{1}1\bar{1}]$) differently oriented NV centers in a [100] diamond crystal.

We have also simulated unpolarized $g^{(2)}(0)$ measurements for two and three-NV center systems as a function of the relative emitter intensities and the background signal. Figure 4 (a) shows simulated unpolarized $g^{(2)}(0)$ measurements of a two-NV center system within a diffraction-limited spot with varying PL intensity ratio between the two NV centers and background. It illustrates the limitations of unpolarized g$^{(2)}$(0) measurements for the identification of the number of NV centers assuming the presence of a finite background and NV centers with variable PL intensities. The presence of unequal intrinsic PL intensity between single-photon emitters can exist due to varying excitation conditions, different alignments, or different local environments such as local nitrogen concentration [25]. g$^{(2)}$(τ) for several excitation polarization angles is shown in Supplementary Material Fig. S4 and demonstrates the excitation polarization dependence.

Figure 4(c) shows the simulated unpolarized g$^{(2)}$(0) value for 3 NV centers present within a diffraction-limited spot without background. The region below the black dashed line shows the second-order correlation measurement g$^{(2)}$(0) value less than 0.5 in the 3 NV center case. This region highlights the importance of performing polarization-dependent g$^{(2)}$(0) measurements for characterization of single-photon emitters.

In Figure 4(b) and (d) we show how ODMR can determine the number of emitters for cases of one, two, and three NV centers located within a diffraction-limited spot in an [100] oriented diamond crystal. Simulated ODMR spectra for two and three NV center systems are shown in Figure 5(d). NV orientations $\vec{a}$ and $\vec{c}$ were simulated for the two-NV center system and two $\vec{a}$ and one $\vec{c}$ for the three NV centers system. In both the two and three-NV center systems, we assume an external magnetic field in the $\left[\frac{1}{3}\frac{1}{3}1\right]$ direction. The two resonance frequencies close to the zero-field magnetic resonance frequency of 2.857 GHz display half of the intensity contrast in theoretical simulation compared to the outer two resonance frequencies. In principle, this allows the experimental distinction of these two cases. However, in experiments, this distinction can be challenging due to experimental noise (see Supplemental Material Figure S9 [17]), and may require pulse microwave sequencing, high precision lock-in system and more complicated techniques. Our research focuses on an all-optical solution for determination of collocated fluorescence emitters independent of magnetic field spin measurements.

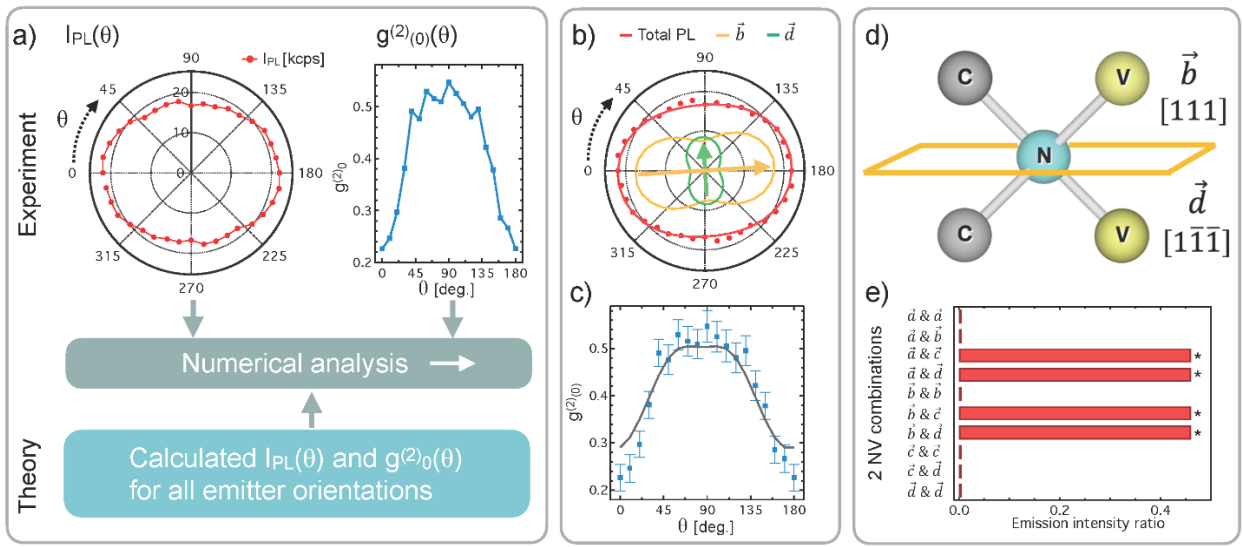

**Figure 5.** (a) Flowchart of the analysis process for the identification of the number and orientation of SPEs. Experimental data and simulations of the PL intensity and g$^{(2)}$(0) as a function of emission polarization angle. Experimental and theoretical results are analyzed using a $\chi^2$ optimization protocol. The results of the numerical analysis for PL intensity and g$^{(2)}$(0) as a function of emission polarization angle are shown in (b) and (c), respectively. (d) Physical representation of the analysis result displaying the vector coordinates of the resolved two NV centers. (e) Result of intrinsic PL intensity ratio (red bar) and orientations of the two NV centers corresponding to minimum $\chi^2$ value highlighted with (*) identified using the $\chi^2$ optimization protocol. The orientations configurations $\vec{a}\&\vec{c}$, $\vec{a}\&\vec{d}$, $\vec{b}\&\vec{c}$, and $\vec{b}\&\vec{d}$ have the same value of intrinsic PL intensity ratio and $\chi^2$ value, indicating that this point has an equal probability of being in any of these configurations. Note that these configurations are degenerate with respect to the experimental orientation and are therefore indistinguishable.

**Experimental results**

We analyze experimental results using our developed protocol on several fluorescent regions on the sample (see Supplemental Material Fig. S8 [17]). These regions all showed a g$^{(2)}$(0) value between 0.4 and 0.6 in the absence of an emission polarizer and were chosen because they may contain one or two NV centers. Measurements of PL intensity and g$^{(2)}$(0) values as a function of emission polarization angle for a spot with an unpolarized g$^{(2)}$(0) value of 0.454 ± 0.018 are shown in Fig. 5(b) and 5(c), respectively. g$^{(2)}$(0) was varied from 0.22 to 0.52 (Fig. 5(c)) as the emission polarization angle rotated from 0 to 90 deg, and for the first time clearly demonstrates a significant dependence of g$^{(2)}$(0) on emission polarization angle. Furthermore, it also proves that more than one NV center must be located within the investigated PL spot.

NV orientations configurations $\vec{a}\&\vec{c}$, $\vec{a}\&\vec{d}$, $\vec{b}\&\vec{c}$, and $\vec{b}\&\vec{d}$ show a $\chi^2$ value of < 0.1, while all other combinations yield values above 0.35 (see Fig. 5e). Hence, we have identified NV centers with orientation $\vec{a}$ and $\vec{c}$ (and their undistinguishable equivalents) as the origin of the observed PL.

The PL intensity ratio was also calculated for each combination of two NV centers (Fig. 5e). The accuracy of the determined PL intensity ratio and background was verified using a confidence interval variance analysis of the results. While the NV orientation combinations identified via the above $\chi^2$ analysis (i.e. $\vec{a}\&\vec{c}$, $\vec{a}\&\vec{d}$, $\vec{b}\&\vec{c}$, and $\vec{b}\&\vec{d}$) yield a PL intensity ratio of 0.46 (e.g. PL intensity of $\vec{a}/\vec{c}$ is 0.46), all others show PL intensity ratios below 1×10$^{-6}$. Such low-intensity ratios represent a high single-emitter PL intensity dominating results. All intensity ratios below 1×10$^{-6}$ also yield $\chi^2$ values above 0.35, which then indicates a poor fit and demonstrates that the experimental data cannot be explained by the presence of only one emitter. Therefore, our protocol has successfully identified $\vec{a}\&\vec{c}$ (and their undistinguishable equivalents) with a PL intensity ratio $\vec{a}/\vec{c}$ of 0.46 as the two single photon emitters present in the investigated PL region. Importantly, based on the unpolarized g$^{(2)}$(0) value of 0.454 ± 0.018 of these two-emitter regions and the commonly used g$^{(2)}$(0) threshold value of 0.5, these signals might have been incorrectly identified as arising from one single photon emitter.

## Conclusion

In conclusion, we have developed and demonstrated a new all-optical method for the determination of one or two emitter systems using a single linear polarizer and the method also provides the characterization of the orientation and relative PL intensity in a two-NV center system. Furthermore, we have shown that the combination of polarization and quantum correlation measurement can provide much information on the NV center color defect.

Our protocol can offer precise knowledge of defects for quantum computing technologies, which affects the precision of resulting quantum protocols. This research provides a valuable new all-optical mechanism for the determination of one or two emitter systems useful for quantum sensing, communication, and computation tasks.

## Acknowledgments

D.P. would like to acknowledge the support of a Ph.D. scholarship from the ARC Centre of Excellence for Nanoscale BioPhotonics and RMIT University. This work was supported with funding from the Air Force Office of Scientific Research (FA9550-20-1-0276) and the Australian Research Council (CE140100003 and FT160100357). P.R. acknowledges support through an Australian Research Council DECRA Fellowship (grant no. DE200100279) and an RMIT University Vice-Chancellor's Research Fellowship. Part of this study was carried out within the framework of the QST International Research Initiative.

# Supplemental Material:

# All-optical determination of one or two emitters using quantum polarization with nitrogen-vacancy centers in diamond

Davin Yue Ming Peng,[1] Josef G. Worboys,[1] Qiang Sun,[1] Shuo Li,[1] Marco Capelli,[1] Shinobu Onoda,[2] Takeshi Ohshima,[2] Philipp Reineck,[1] Brant C. Gibson,[1] and Andrew D. Greentree[1]

[1]*Australian Research Council Centre of Excellence for Nanoscale Biophotonics, RMIT University, Melbourne, VIC 3001, Australia*

[2]*Takasaki Advanced Radiation Research Institute, National Institutes for Quantum Science and Technology,1233 Watanuki, Takasaki, 370-1292, Gunma, Japan*

## I. Derivation of the second-order correlation formula for two NVs

Photoluminescence (PL) from nitrogen-vacancy (NV) centers is properly treated as arising from either of two orthogonal dipoles. These transitions are energetically degenerate but lead to different polarization properties, hence will have different contributions to the PL signal when monitored through a polarizer. A schematic showing the double-dipole model that we use here is shown in Fig. S1. NV PL is modeled as occurring via one of the two dipoles, or more generally via a superposition of both dipoles, with linear electrical dipole x and dipole y perpendicular to each other and where the z-axis is defined along the NV axis. The probability of detecting photons from dipole $i = x,y$ from $NV_k$ $k$ = a,b is $P_{i,k}$, which is in general a function of the angle of the polarizer.

We are concerned with the problem of detecting the Hanbury Brown and Twiss (HBT) correlations from a two-NV system. Taking into account the fact that a given NV cannot de-excite simultaneously via both dipoles and assuming no coupling between the NV centers, then we may write down the expected g2(0) for the two-emitter system as

$$g^{(2)}(0) = \frac{2(P_{x,a}P_{x,b}+P_{x,a}P_{y,b}+P_{x,b}P_{y,a}+P_{y,b}P_{y,a})}{(P_{x,a}+P_{y,a}+P_{x,b}+P_{y,b})^2}.$$

*(S1)*

By summing the probabilities associated with the two de-excitation pathways with $P_k = P_{x,k} + P_{y,k}$, then Eq. S1 reduces to Eq.3 from the main text.

Following the approach in [2] we can also obtain $g^{(2)}$ with background under the double dipole assumption:

$$g^{(2)}(0) = \frac{2(P_{x,a} + P_{y,a})(P_{x,b} + P_{y,b}) + 2(P_{x,a} + P_{y,a} + P_{x,b} + P_{y,b})NP_\gamma + (NP_\gamma)^2}{(P_{x,a} + P_{y,a} + P_{x,b} + P_{y,b} + NP_\gamma)^2}$$

*(S2)*

where we have treated the background noise as arising from a large bath of $N \gg 1$ emitters, with low detection probability $P_\gamma \ll P_{x,k}, P_{y,k}$ where $NP_\gamma$ is not negligible.

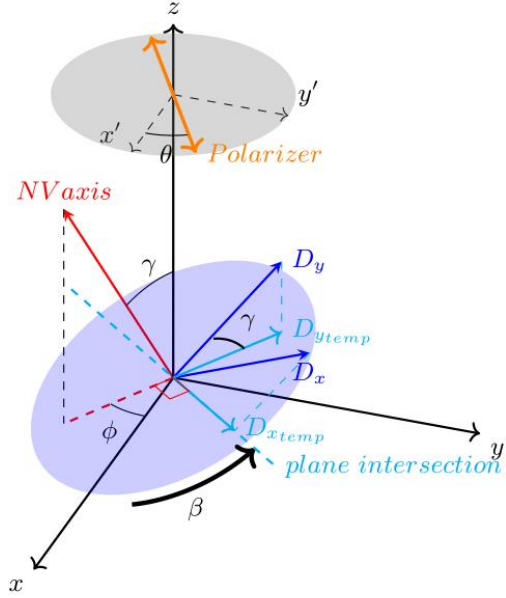

**Fig. S1.** Schematic of the double-dipole ($D_x$, $D_y$) emission model of an NV center (red arrow), including the detection polarizer (yellow arrow and grey oval). There are two important reference frames here. Firstly, the z-axis is defined by the optical axis of the system and is orthogonal to the diamond surface. The second reference frame is that of the NV axis (red arrow), which is defined by the location of nitrogen and vacancy atoms in the diamond crystal lattice. The two NV emission dipoles ($D_x$, $D_y$) are orthogonal to the NV axis and each other which defines the NV x-y plane.

Because the emission polarization from each dipole is different, each dipole pathway contributes different emission intensities through the linear polarizer. With the angle definitions from Fig. S1, we may write down the polarization and orientation-dependent result for *g₂(0)* which is

$$g^{(2)}(0)(\theta, \beta_a, \beta_b, \gamma_a, \gamma_b, \phi_a, \phi_b)$$
$$= \frac{\begin{array}{l} 2\{\cos^2\theta\,(\sin^2\phi_a + \cos^2\gamma_a \cos^2\phi_a) + \sin^2\theta\,(\cos^2\phi_a + \cos^2\gamma_a \sin^2\phi_a)\} \\ \{+\cos^2\theta\,(\sin^2\phi_b + \cos^2\gamma_b \cos^2\phi_b) + \sin^2\theta\,(\cos^2\phi_b + \cos^2\gamma_b \sin^2\phi_b)\} \\ +2\left\{\begin{bmatrix} \cos^2\theta\,(\sin^2\phi_a + \cos^2\gamma_a \cos^2\phi_a) + \sin^2\theta\,(\cos^2\phi_a + \cos^2\gamma_a \sin^2\phi_a) \\ +\cos^2\theta\,(\sin^2\phi_b + \cos^2\gamma_b \cos^2\phi_b) + \sin^2\theta\,(\cos^2\phi_b + \cos^2\gamma_b \sin^2\phi_b) + NP_\gamma \end{bmatrix}^2\right\} NP_\gamma + (NP_\gamma)^2 \end{array}}{\begin{bmatrix} \cos^2\theta\,(\sin^2\phi_a + \cos^2\gamma_a \cos^2\phi_a) + \sin^2\theta\,(\cos^2\phi_a + \cos^2\gamma_a \sin^2\phi_a) \\ +\cos^2\theta\,(\sin^2\phi_b + \cos^2\gamma_b \cos^2\phi_b) + \sin^2\theta\,(\cos^2\phi_b + \cos^2\gamma_b \sin^2\phi_b) + NP_\gamma \end{bmatrix}^2}$$

*(S3)*

The probability of detecting photons from one NV center ($P_{NV}^{\text{detect}}$) can be expressed as the sum of two NV dipole's photon detecting probability:

$$P_{NV}^{\text{detect}} = P_{x,k} + P_{y,k} = \cos^2\theta\,(\sin^2\phi + \cos^2\gamma\cos^2\phi) + \sin^2\theta\,(\cos^2\phi + \cos^2\gamma\sin^2\phi)$$

(S4)

with the double-dipole rotation-angle $\beta$ canceled out. As such, the $g^{(2)}$ formulation based on the double-dipole assumption is identical to the formulation that assumes a single dipole.

The photon detection rate D of a single dipole emitter in experimental setup with linear polarizer filter can therefore be expressed as a surface integral of the collection plane.

$$D_n(\theta) = \int \left| E_{x'}\cos(\theta) + E_{y'}\sin(\theta) \right|^2 dS$$

(S5)

Where $E_{x'}$ and $E_{y'}$ are the x and y component of the electric field expression $\boldsymbol{E}$ of a single dipole emitter when applying a linear polarizer filter

$$\boldsymbol{E} = \frac{1}{4\pi\epsilon_0\epsilon_r}\left\{\frac{k^2}{r^3}(\boldsymbol{r}\times\boldsymbol{p})\times\boldsymbol{r} + \left(\frac{1}{r^5} - \frac{ik}{r^4}\right)[3\boldsymbol{r}(\boldsymbol{r}\cdot\boldsymbol{p}) - r^2\boldsymbol{p}]\right\}e^{ikr}$$

(S6)

where $i = \sqrt{-1}$, $\boldsymbol{p}$ is the electric dipole moment to represent the NV center at $\boldsymbol{x}_0$, $\boldsymbol{r} = \boldsymbol{x} - \boldsymbol{x}_0$ with $\boldsymbol{x}$ being the location of the collector element, $r = |\boldsymbol{x} - \boldsymbol{x}_0|$, $\epsilon_0$ is the vacuum permittivity, $\epsilon_r$ is the relative permittivity and $k$ is the wavenumber as $k\lambda = 2\pi$.

## II. Simulation of the emission intensity of the NV double-dipole as a function of the emission polarization angle

Figure. S2 shows the simulated emission intensity of all four possible NV double-dipole orientations in an [100] oriented single crystal diamond for the orientation angle $\beta = 0°$ (see Fig. S1) as a function of emission polarization angle (polar axis). The blue and red traces show the emission intensities ($PL_{NV,x,k}$, $PL_{NV,y,k}$, radial axis) of the two individual dipoles in arbitrary units. The black traces represent the sum of the two individual dipoles. Importantly, the sum of the two dipoles remains constant with respect to varying dipole orientation angle $\beta$ (not shown in the figure). The latter is illustrated in the Supplementary Material video "*Two dipole emission intensity simulation with variation of dipole orientation angle.MP4*" which shows the evolution of all emission dipole intensities as a function of dipole orientation angle $\beta$. This confirms that although a single NV center consists of two orthogonal emission dipoles, its total emission is independent of their orientation angle $\beta$ (see also Eq. S4)

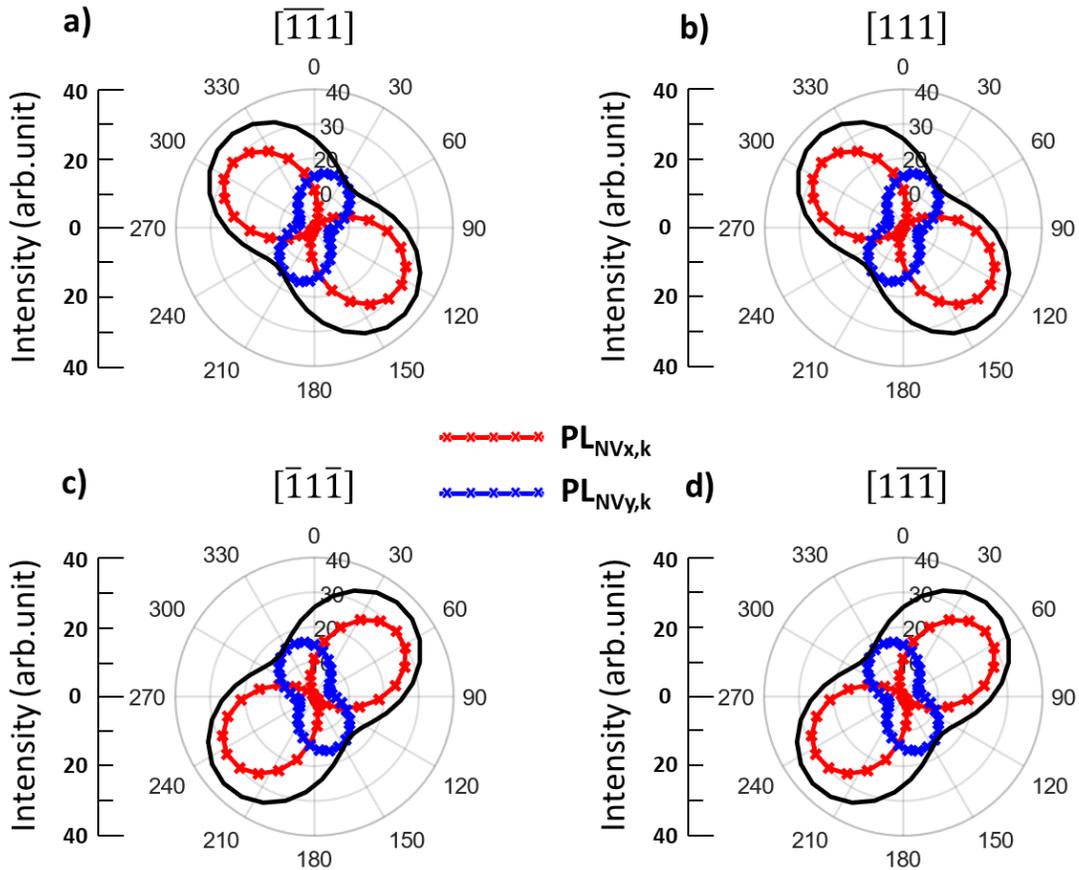

**Fig. S2.** Simulation of the emission intensity (in arbitrary units) from two orthogonal dipole moments (blue and red traces) of all four NV center orientations (a - d) in an [100] oriented single crystal diamond with dipole orientation angle $\beta = 0°$. The black trace shows the sum of the two dipole emission intensities.

## III. The effect of excitation polarization on NV single photon emission properties

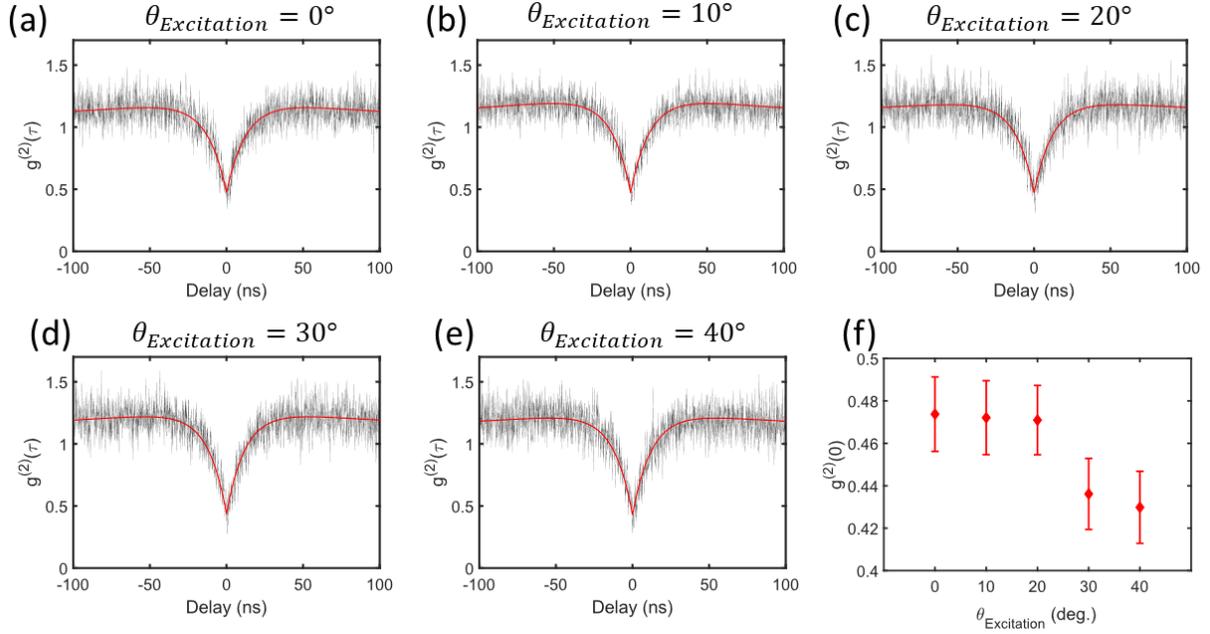

**Fig. S3.** (a) - (e) Second order correlation $g^{(2)}$ measurements for different excitation beam polarizations ($\theta_{\text{excitation}}$) as indicated above each graph. A 532 nm CW laser (500 µW total beam power) was used for excitation. All emitted photons were collected – irrespective of their polarization. (f) $g^{(2)}(0)$ values of the autocorrelation functions plotted in (a) to (e) as a function of excitation beam polarization. All measurements were performed on the two NV single photon emitters investigated in Fig. 2 in the main text. The change in $g^{(2)}(0)$ as a function of excitation beam polarization is caused by the preferential excitation of one of the NV centers within the focal point of the beam [1], causing a change in the relative brightness of the two NV centers. [2]

## IV. Experimental setup

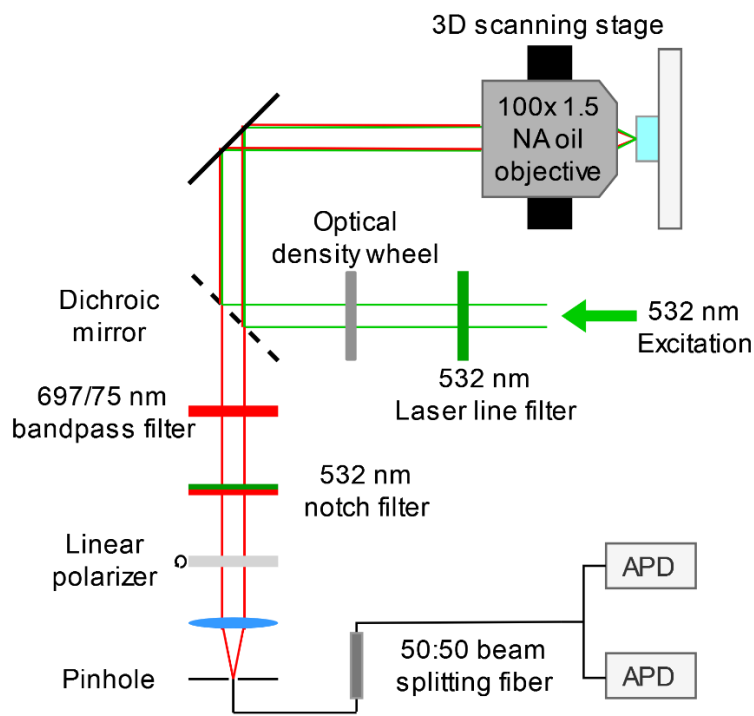

**Fig. S4.** Illustration of the experimental setup. A collimated laser beam (532 nm, ~500 µW total beam power) was passed through a laser line filter (LL01-532 , Semrock, USA) and focused into the sample with a 100x oil immersion objective (NA 1.5, UPLAPO100XOHR, company, country Japan) mounted on a 3D scanning stage (P-545.3R7, Physik Instrumente, Germany). To collect a PL image, the objective was raster scanned across the sample. PL was collected with the same objective and separated from the excitation signal using a dichroic (Di02-R532-25x36, Semrock, USA) a 532 nm notch filter (NF01-532U-25, Semrock, USA), and a bandpass filter (FF01-697/75-25-D, Semrock, USA). It was then passed through a linear polarizer, fiber coupled into a 50:50 beam-splitting fiber, and detected using avalanche photodiodes (SPCM-AQRH-14, Excelitas). Photon incidences were analyzed using a correlator card (TimeHarp260, Picoquant GmBH, Germany).

## V. PL intensity analysis of adenine ($C_5H_5N_5$) implanted electronic grade CVD diamond sample

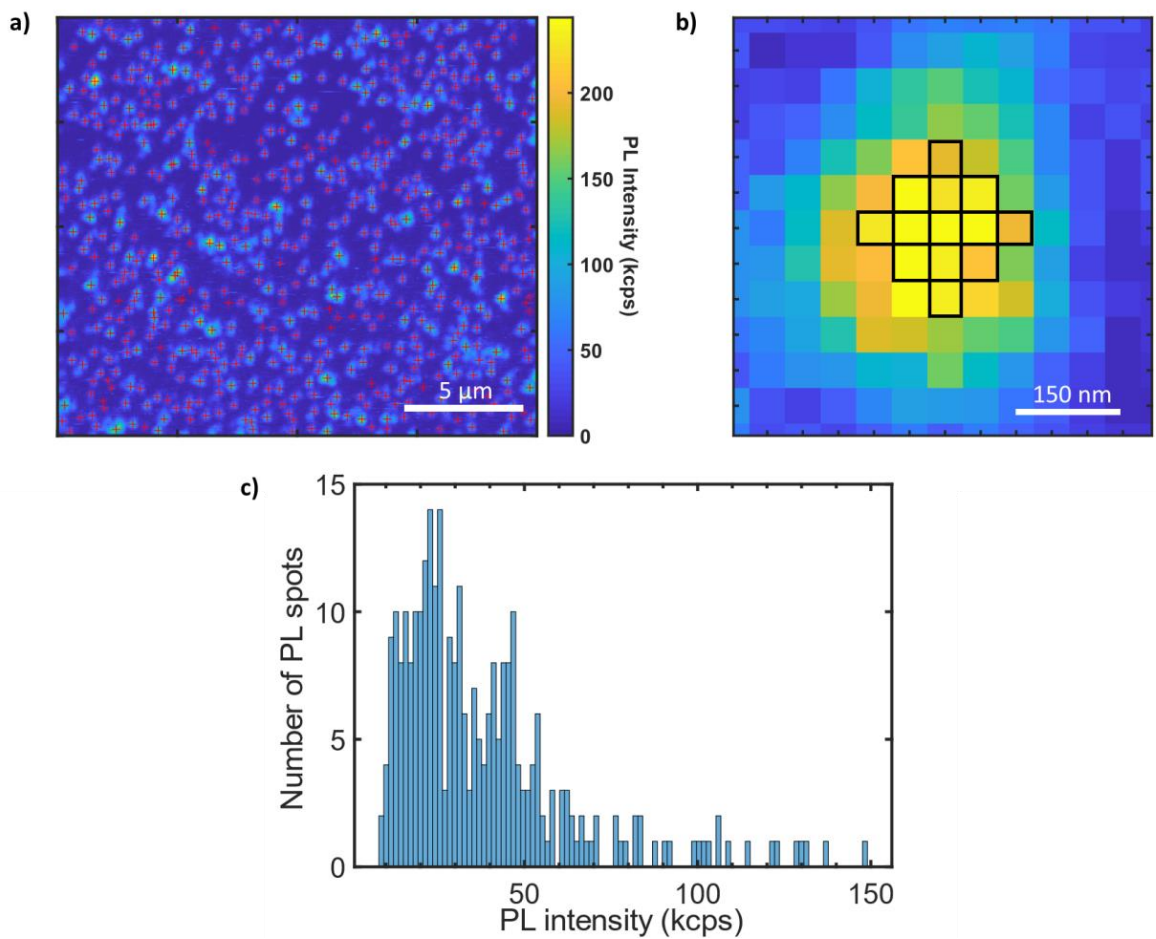

**Fig. S5.** a) PL intensity map of a 20 x 20 µm region on an adenine ($C_5H_5N5_5$) implanted electronic grade CVD diamond sample. A 532 nm continuous wave laser (500 µW) was used for excitation and the PL image was acquired as described in Fig S3. A "*FastPeakFind.m*" (MATLAB) script was used to identify regions of interest that likely contain one or more single photon emitters. A threshold value of 5000 counts per second (cps) was used and the 362 identified spots are indicated by red markers in the PL intensity map. For each spot, the PL intensity was averaged over a two-pixel radius around the pixel containing the peak as illustrated in panel b). c) shows the PL intensity histogram for all locations marked in panel a). The histogram shows a broad distribution of PL intensities, with only one identifiable peak. This demonstrates that individual emitters have different PL intensities and that an unambiguous identification of the number of emitters based on brightness is not feasible.

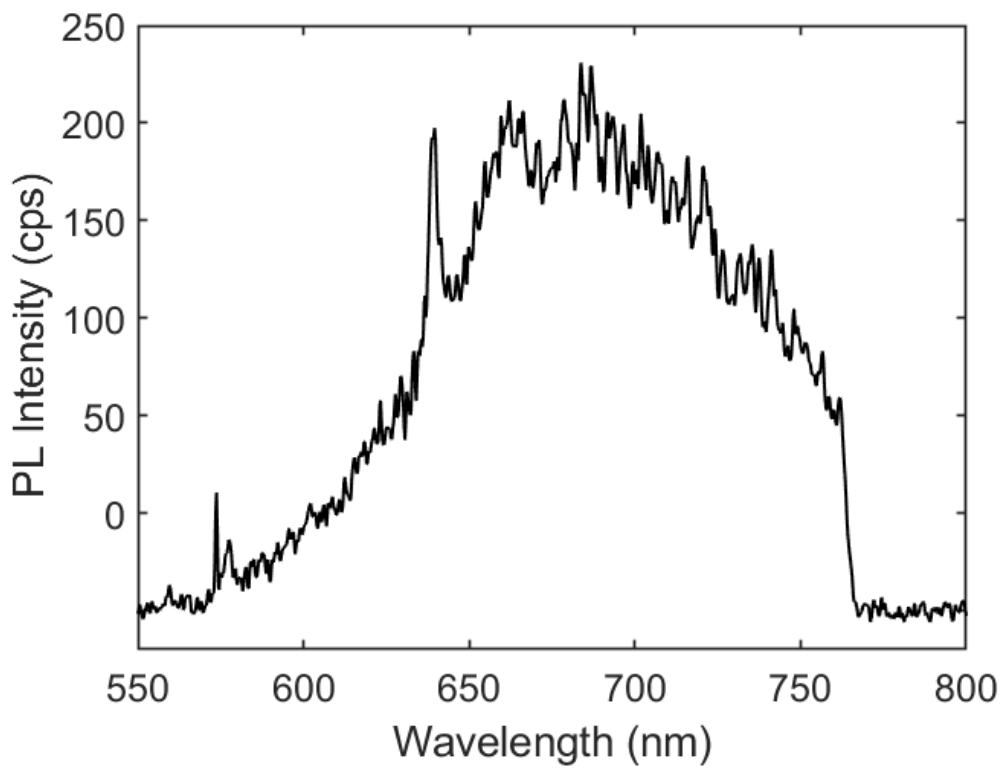

**Fig. S6.** PL spectrum of a fluorescence spot on the adenine (C5H5N55) implanted electronic grade CVD diamond sample under 532 nm continuous wave laser (500 uW) excitation and 532 nm notch and 697/75 nm filter collection conditions bandpass at room temperature. Spectra information identifies the presence of both NV$^0$ and NV$^-$ indicated by peaks at 575 nm and 638 nm.

## VI. The effect of a background signal on the developed protocol

To investigate the effect of a background signal on the accuracy of the $\chi^2$ optimization protocol, the absolute error between simulated data and optimization results was investigated as a function of the background intensity. We investigated the error between simulated data and the results obtained by applying the $\chi^2$ optimization protocol to the simulated data. Figure 5 shows the results for the error in PL intensity ratio between the two emitters (Figure 5 (a)), the background error (Figure 4 (b)) and $\chi^2$ (Figure 5 (c)) for a wide range of background signals (x-axis) and PL intensity ratios (y-axis). The dashed vertical red line indicates the background signal that corresponds to 20% of the intensity of NV1. The solid red traces in Figure 5 (a) and (b) are the contour traces of a PL intensity ratio error of 0.1 and background error of 5, respectively.

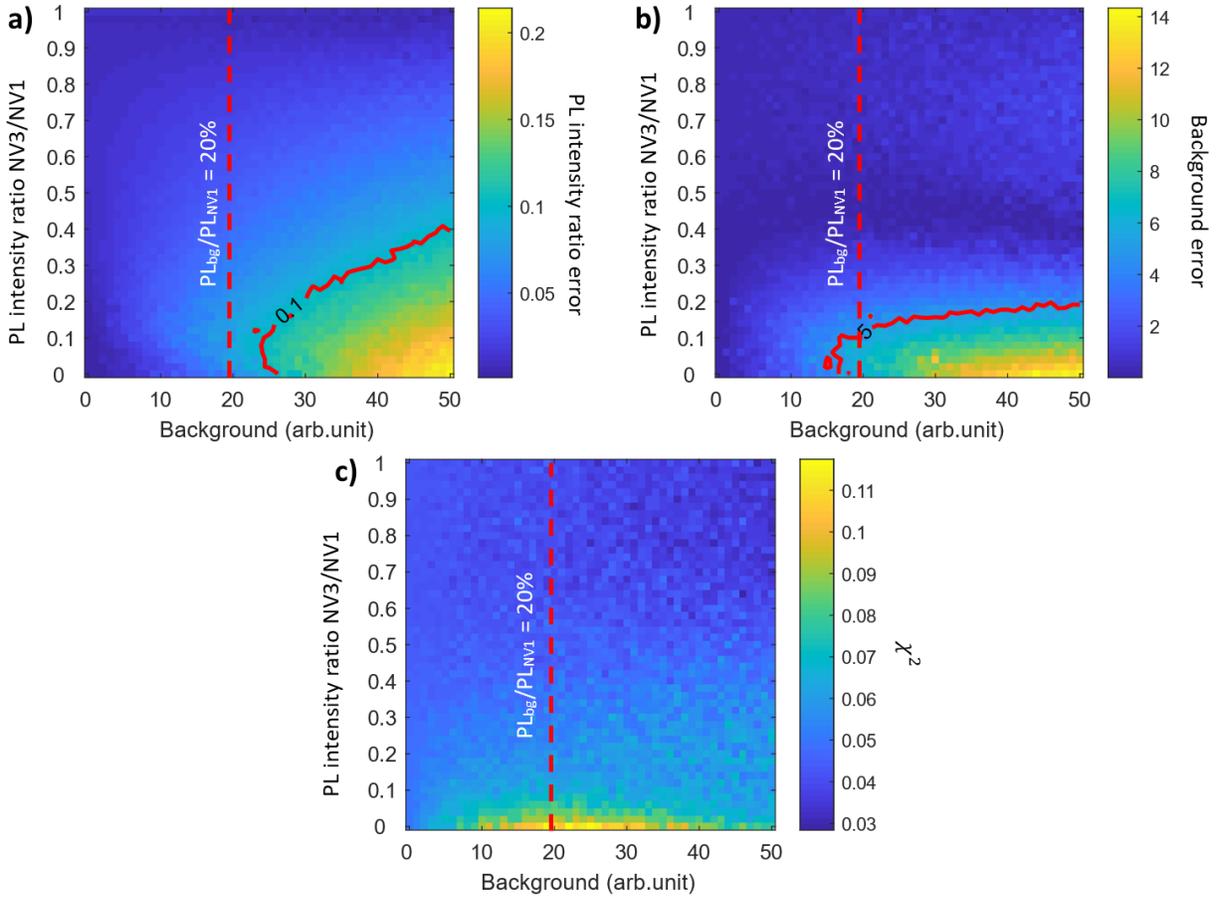

**Figure S7.** The effect of a background signal on the developed optimization protocol for a two-NV center system ($\vec{a}$ and $\vec{c}$) a) PL intensity ratio error (color scale) as a function of background intensity and PL intensity ratio. b) Background intensity error as a function of background intensity and PL intensity ratio. c) $\chi^2$ as a function of background intensity and PL intensity ratio.

## VII. Analysis of five different PL regions

In this section, we analyze five different PL regions using the developed protocol. Fig. S10 shows the main results of the $\chi^2$ optimization protocol for five PL regions (I to V) that exhibit a $g^{(2)}(0)$ in the range between 0.4 and 0.6 for some emission polarization angles. In a) and b), the normalized intensity and $g^{(2)}(0)$ are plotted on the radial axis, respectively, and the emission polarization.

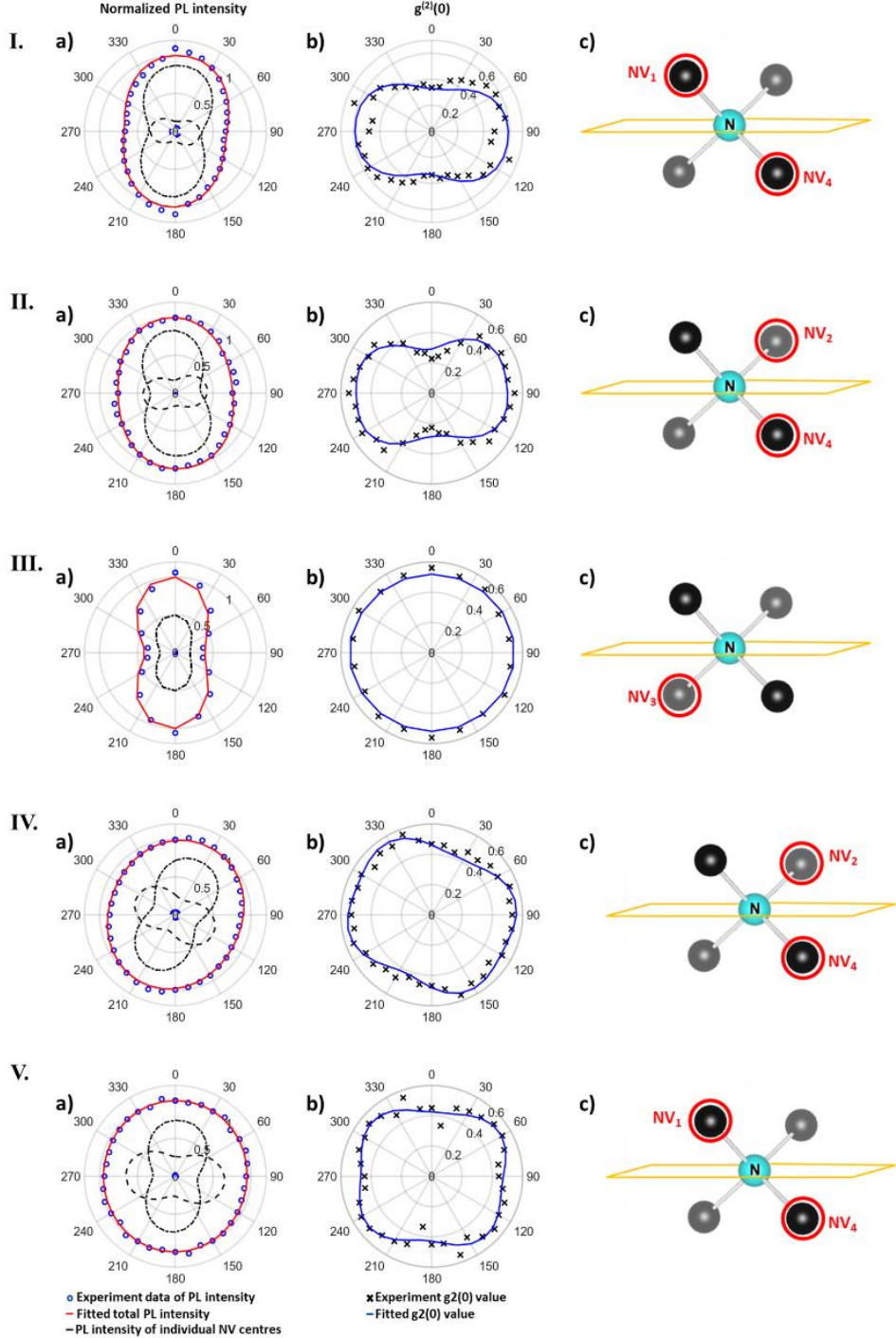

**Fig. S8.** Main results of the $\chi^2$ optimization protocol for five PL regions.

## VIII. Simulation of ODMR spectra with and without noise

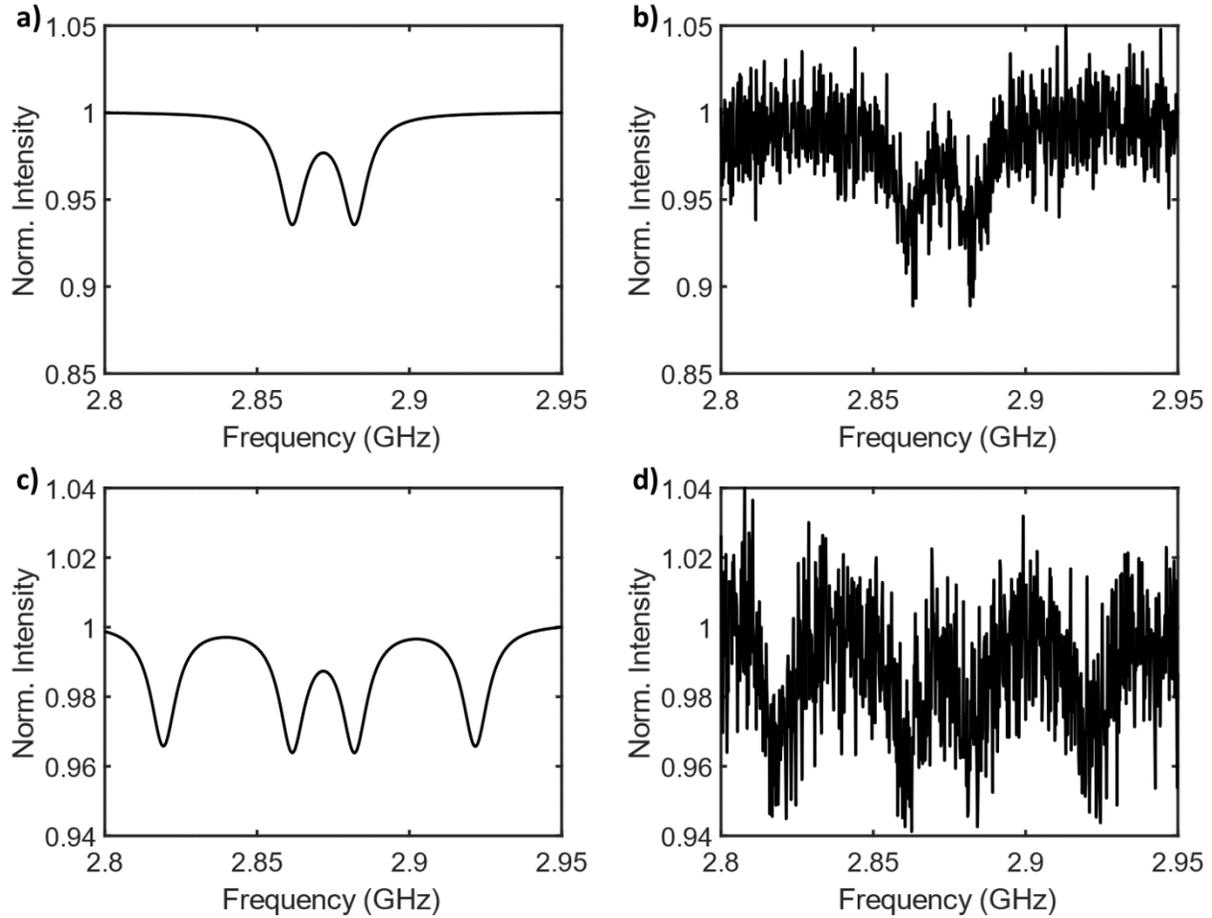

**Fig. S9.** The effect of noise on simulated ODMR spectra of a single (a and b) and two-NV center system (c and d). In c) and d) the two NV centers have different orientations in an [100] oriented diamond crystal. a) and c) show the simulated normalized PL intensity as a function of microwave frequency without noise. In b) and d) Poisson noise was added to the spectra shown in a) and c) to illustrate that the unambiguous identification of the number of emitters using ODMR is non-trivial even for idealized simple 2 NV model systems, where both emitters have the same brightness and different orientations.